\documentclass{PoS}
\usepackage{braket}
\usepackage{amsmath}
\usepackage{amsbsy}
\usepackage{amsfonts}
\usepackage{amssymb}
\usepackage{graphics}
\usepackage{graphicx}
\usepackage{capt-of}
\usepackage{csquotes}
\usepackage{dcolumn}

\title{Construction and quality assurance of the Belle II Silicon Vertex Detector}

\ShortTitle{Belle II silicon vertex detector}

\author{\speaker{P.~K.~Resmi}$^f$, H.~Aihara$^r$, T.~Aziz$^j$, S.~Bacher$^v$, S.~Bahinipati$^e$, E.~Barberio$^a$, Ti.~Baroncelli$^a$, To.~Baroncelli$^a$, A.~K.~Basith$^f$, G.~Batignani${k,l}$, A.~Bauer$^b$, P.~K.~Behera$^f$, V.~Bertacchi$^{k,l}$, S.~Bettarini$^{k,l}$, B.~Bhuyan$^g$, T.~Bilka$^d$, F.~Bosi$^l$, L.~Bosisio$^{m,n}$, A.~Bozek$^v$, F.~Buchsteiner$^b$, G.~Caria$^a$, G.~Casarosa$^{k,l}$, M.~Ceccanti$^l$, D.~\v{C}ervenkov$^d$, T.~Czank$^q$, N.~Dash$^e$, M.~De~Nuccio$^{k,l}$, Z.~Dole\v{z}al$^d$, F.~Forti$^{k,l}$, M.~Friedl$^b$, B.~Gobbo$^n$, J.~A.~M. Grimaldo$^r$, K.~Hara$^s$, T.~Higuchi$^o$, C.~Irmler$^b$, A.~Ishikawa$^q$, H.~B.~Jeon$^t$, C.~Joo$^o$, M.~Kaleta$^v$, J.~Kandra$^d$, K.~H.~Kang$^t$, P.~Kody\v{s}$^d$, T.~Kohriki$^s$, I.~Komarov$^n$, M.~Kumar$^h$, R.~Kumar$^i$, W.~Kun$^r$, P.~Kvasni\v{c}ka$^d$, C.~La Licata$^{m,n}$, K.~Lalwani$^h$, L.~Lanceri$^{m,n}$, J.~Y.~Lee$^u$, S.~C.~Lee$^t$, Y.~Li$^c$, J.~Libby$^f$, T.~Lueck$^{k,l}$, P.~Mammini$^l$, A.~Martini$^{k,l}$, S.~N.~Mayekar$^j$, G.~B.~Mohanty$^j$, T.~Morii$^o$, K.~R.~Nakamura$^s$, Z.~Natkaniec$^v$, Y.~Onuki$^r$, W.~Ostrowicz$^v$, A.~Paladino$^o$, E.~Paoloni$^{k,l}$, H.~Park$^t$, K.~Prasanth$^j$, A.~Profeti$^l$, K.~K.~Rao$^j$, I.~Rashevskaya$^{n,A}$, G.~Rizzo$^{k,l}$, M.~Rozanska$^v$, D.~Sahoo$^j$, J.~Sasaki$^r$, N.~Sato$^s$, S.~Schultschik$^b$, C.~Schwanda$^b$, J.~Stypula$^v$, J.~Suzuki$^s$, S.~Tanaka$^s$, H.~Tanigawa$^r$, G.~N.~Taylor$^a$, R.~Thalmeier$^b$, T.~Tsuboyama$^s$, P.~Urquijo$^a$, L.~Vitale$^{m,n}$, M.~Watanabe$^{p,B}$, S.~Watanuki$^q$, I.~J.~Watson$^r$, J.~Webb$^a$, J.~Wiechczynski$^v$, S.~Williams$^a$, H.~Yin$^b$, L.~Zani$^{k,l}$
(Belle~II~SVD~Collaboration)

        \\
       $^a$School of Physics, University of Melbourne, Melbourne, Victoria 3010, Australia\\
       $^b$Institute of High Energy Physics, Austrian Academy of Sciences, 1050 Vienna, Austria\\
       $^c$Peking University, Department of Technical Physics, Beijing 100871, China\\
       $^d$Faculty of Mathematics and Physics, Charles University, 121 16 Prague, Czech Republic\\
       $^e$Indian Institute of Technology Bhubaneswar, Satya Nagar, India\\
       $^f$Indian Institute of Technology Madras, Chennai 600036, India\\
       $^g$Indian Institute of Technology Guwahati, Assam 781039, India\\
       $^h$Malviya National Institute of Technology, Jaipur 302017, India\\
       $^i$Punjab Agricultural University, Ludhiana 141004, India\\
       $^j$Tata Institute of Fundamental Research, Mumbai 400005, India\\
       $^k$Dipartimento di Fisica, Universit\`{a} di Pisa, I-56127 Pisa, Italy\\
       $^l$INFN Sezione di Pisa, I-56127 Pisa, Italy\\
       $^m$Dipartimento di Fisica, Universit\`{a} di Trieste, I-34127 Trieste, Italy\\
       $^n$INFN Sezione di Trieste, I-34127 Trieste, Italy, $^A$Presently at TIFPA-INFN, Dipartimento di Fisica, Universit\`{a} di Trento, I-38123 Trento, Italy\\
       $^o$Kavli Institute for the Physics and Mathematics of the Universe (WPI), University of Tokyo, Kashiwa 277-8583, Japan\\
       $^p$Department of Physics, Niigata University, Niigata 950-2181, Japan, $^B$Presently at Nippon Dental University, Niigata 951-8580, Japan\\
       $^q$Department of Physics, Tohoku University, Sendai 980-8578, Japan\\
       $^r$Department of Physics, University of Tokyo, Tokyo 113-0033, Japan\\
       $^s$High Energy Accelerator Research Organization (KEK), Tsukuba 305-0801, Japan\\
       $^t$Department of Physics, Kyungpook National University, Daegu 702-701, Korea\\
       $^u$Department of Physics and Astronomy, Seoul National University, Seoul 151-742, Korea\\
       $^v$H. Niewodniczanski Institute of Nuclear Physics, Krakow 31-342, Poland
       
       E-mail: \email{resmipk@physics.iitm.ac.in}}
       
\abstract{The Belle II experiment, which is situated at the interaction point of the SuperKEKB $e^+e^{-}$ collider at KEK, Tsukuba, Japan, is expected to collect data corresponding to an integrated luminosity of 50~ab$^{- 1}$. This data set will be sensitive to beyond-the-standard-model physics via precision measurements and searches for very rare decays. At its heart lies a six-layer vertex detector consisting of two layers of pixel detectors (PXD) and four layers of double-sided silicon microstrip detectors (SVD). Precise vertexing as provided by this device is essential for measurements of time-dependent $CP$ violation. Crucial aspects of the SVD assembly are precise alignment, as well as rigorous electrical and geometrical quality assurance. We present an overview of the construction of the SVD, including the precision gluing of SVD component modules and the wire-bonding of various electrical components. We also discuss the electrical and geometrical quality assurance tests.}

\FullConference{ The 27th International Workshop on Vertex Detectors\\
         21-26 October 2018\\
         MGM Beach Resorts, Chennai, India}
       
\begin{document}

\section{Introduction}
The Belle II~\cite{belle2} experiment is a next generation flavour factory with physics goals that include, but not limited to, studies of $CP$ violation in $B$ and $D$ meson sectors, searches for rare decays, tests of lepton flavour (universality) violation and the search for dark sector particles and forces. These physics goals are indirect probes of physics beyond the standard model. It is an upgraded version of its predecessor, Belle~\cite{belle}. The Belle II detector sits at the interaction point of the SuperKEKB asymmetric $e^{+}e^{-}$ collider~\cite{superkekb} at KEK, Tsukuba, Japan. It is expected to collect data corresponding to an integrated luminosity of 50 ab$^{-1}$. The target peak luminosity of SuperKEKB is 8~$\times$~10$^{35}$~cm$^{-2}$s$^{-1}$. The beam current will be twice and beam size $\frac{1}{20}$ as that of KEKB.

The inner tracking system consists of two layers of DEPFET based pixel detectors (PXD) and a four layer silicon vertex detector (SVD), which consists of double-sided silicon microstrip sensors. This improves the vertex resolution significantly compared to Belle, which is crucial for measurements of time-dependent $CP$ violation. The impact parameter resolution is 20 $\mu$m at 2 GeV/$c$~\cite{belle2}, which is almost half that of Belle. The reconstruction efficiency of low momentum particles and relatively long-lived particles like $K_{\rm S}^0$ will be improved due to the larger radius of the outermost SVD layer. The physics run with the full detector, except for one layer of PXD, is expected to start in early 2019.

\section{Belle II Silicon Vertex Detector}

The Belle II SVD has four layers and its construction is a global effort involving groups from Asia, Australia and Europe. Layer 3 (L3) is constructed by a research group at the University of Melbourne. Layers 4, 5 and 6 (L4, L5 and L6) are built by TIFR India, HEPHY Vienna and Kavli IPMU Japan, respectively. The forward and backward modules for L4, L5 and L6 are provided by INFN Pisa.

\begin{figure}[ht!]
\centering
\includegraphics[width=6cm, height=5cm]{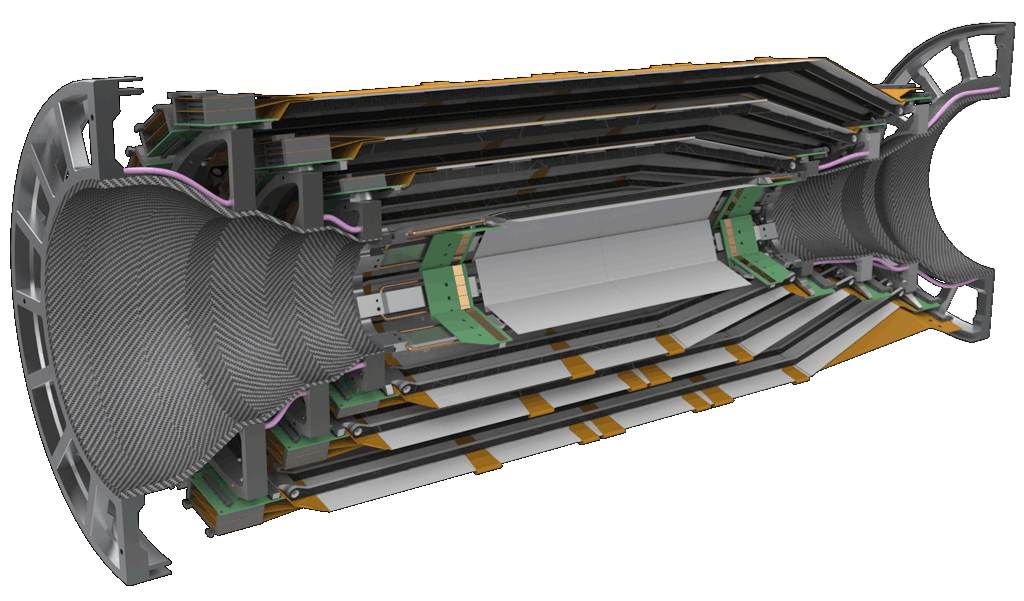}
\caption{A cross-section view of the Belle II SVD. }\label{Fig:svd}
\end{figure}
A cross-section view of the SVD is shown in Fig.~\ref{Fig:svd}. L3, L4, L5 and L6 are made up of seven, 10, 12 and 16 modules (also referred to as "ladders"), respectively. The angular acceptance is $17^{\circ}<\theta < 150^{\circ}$. The innermost layer (L3) has a radius of 39 mm and the outermost one (L6) is 135 mm in radius. The radii for L4 and L5 are 80 mm and 104 mm, respectively. In comparison, the radius of the outermost SVD layer at Belle was 88 mm~\cite{bellesvd2}. The whole structure has a lantern shape to complement the forward boost of the centre of mass system because of the asymmetric beam energies (4 GeV $e^+$ and 7 GeV $e^{-}$). To facilitate this, L4, L5 and L6 have slant angles of 11.9$^{\circ}$, 17.2$^{\circ}$ and 21.1$^{\circ}$, respectively. This structure reduces the material budget without affecting the performance.

\section{Components used for SVD }

\begin{figure}
\centering
\begin{tabular}{cc}
\includegraphics[width=0.33\columnwidth]{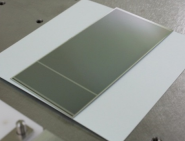}&
\includegraphics[width=0.295\columnwidth]{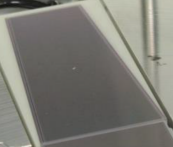}\\[0.5ex]
\end{tabular}
\caption{Rectangular (left) and trapezoidal (right) sensors.}\label{Fig:sensor}
\end{figure}

Three types of double-sided silicon microstrip detectors (DSSD), which are p-in-n type, are used to build the ladders of the SVD. They are six inches long with differing width and have thickness 300 or 320 $\mu$m. The small rectangular DSSDs are used in L3 whereas large rectangular ones are used in L4, L5 and L6. The forward slanted part is built using trapezoidal DSSDs with varying width. The DSSDs are shown in Fig.~\ref{Fig:sensor} and their specifications are given in table~\ref{Table:dssd}.

\begin{table} [t] 
\centering  
 \begin{tabular} { c c c c}
\hline 
 Sensor & Rectangular & Rectangular& Trapezoidal  \\[0.5ex]
& (large) & (small) &  \\[0.5ex]
\hline
\hline
 \# of p-strips & 768 & 768 & 768\\[0.5ex]
 p-strip pitch  & ~75 & ~50 & ~50\textendash75\\[0.5ex]
 ($\mu$m) & & & \\[0.5ex]
 \# of n-strips & 512 & 768 & 512 \\[0.5ex]
 n-strip pitch  & 240  & 160 & 240 \\[0.5ex]
 ($\mu$m) & & & \\[0.5ex]
\hline
\end{tabular}  
\caption{DSSD specifications used in SVD.}\label{Table:dssd}
\end{table} 

The p-side strips are aligned parallel to the beam direction and n-side strips are perpendicular to the beam direction. The L3 DSSDs have their n-side facing the beam pipe whereas L4, L5 and L6 DSSDs are oppositely arranged. This design avoids any interference of L3 support structure with the PXD system. The rectangular DSSDs are manufactured at Hamamatsu Photonics in Japan and the trapezoidal DSSDs are built at Micron Semiconductor in the UK.

\begin{figure}
\centering
\begin{tabular}{cc}
\includegraphics[width=0.25\columnwidth]{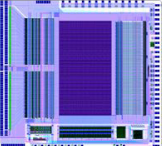}&
\includegraphics[width=0.3\columnwidth]{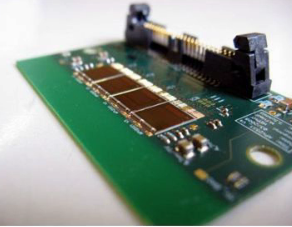}\\[0.5ex]
\end{tabular}
\caption{Image of the APV25 readout chip showing the features and bond pads (left); four APV25 readout chips mounted on a hybrid (right).}\label{Fig:APV}
\end{figure}

The readout chip must have a short signal shaping time in order to cope with the high hit rate expected at Belle II. The APV25 chips~\cite{apv}, originally developed for the CMS Collaboration, is used for this purpose, which has an integration time of 50 ns. It is radiation hard and can tolerate upto 1 MGy, which is far beyond the radiation dose expected at Belle~II (about 50~Gy). It also has a 192 cell deep analog pipeline, which reduces the detector dead-time. The chip is shown in Fig.~\ref{Fig:APV}.

\begin{figure}
\centering
\includegraphics[width=0.6\columnwidth]{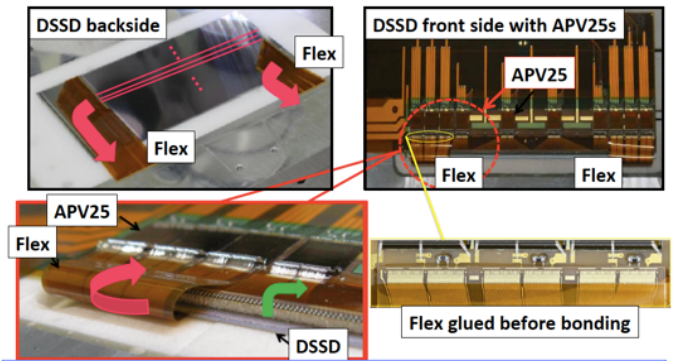}
\caption{Origami chip-on-sensor concept where the readout from the other side of the DSSD is wrapped around using flexible circuits and connected to APV25 chips in a single line on the DSSD.}\label{Fig:origami}

\end{figure}

The APV25 needs to be placed as close to the DSSD as possible to reduce the capacitive noise. This is done with the \enquote{origami} chip-on-sensor concept~\cite{origami}. This novel design allows for the readout chips to be on a single side of the DSSD. The readout channels from the other side are wrapped around via flexible electronic circuits so that the APV25 chips can be placed on a single line (see Fig.~\ref{Fig:origami}). This, in turn, helps in having a single cooling channel, thus reducing the material budget. This origami concept is adopted in the inner DSSDs of L4, L5 and L6. The full L3 ladder and the forward and backward DSSDs are read out from the edges.

A dual-phase CO$_{2}$ cooling system at $-$20$^{\circ}$C is employed to deal with the heat dissipated, approximately 700 W, from all the APV25 chips.

\section{Construction}

\begin{figure}
\centering
\includegraphics[width=0.5\columnwidth]{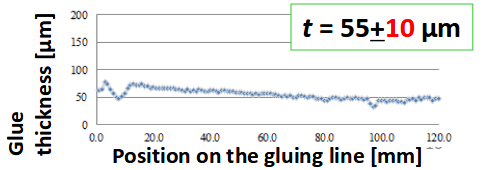}
\caption{Thickness of the glue dispensed by the robotic arm.}\label{Fig:glue}
\end{figure}

The ladder assembly procedure~\cite{ladder} is complex because the DSSDs are aligned precisely using assembly jigs. Vacuum chucking is used to fix the sensors to the jigs. There are different jigs used for various purposes during the entire assembly of one ladder. The flexible circuits are glued to sensor and the electrical connections are made via wire-bonding. Araldite{\textcircled{\raisebox {-0.9pt}{R}}}~ 2011 glue is used and the dispensing is controlled by robotic arm. A uniform glue thickness is achieved with the robotic system as demonstrated in Fig.~\ref{Fig:glue}.

\begin{figure}
\centering
\includegraphics[width=0.5\columnwidth]{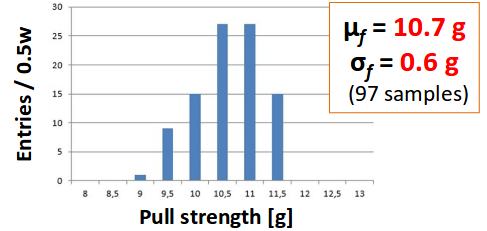}
\caption{Pull strength measured for 97 samples of wire-bonds.}\label{Fig:pull}
\end{figure}

The wire-bonding machine uses Aluminuim wire for the connections. The machine parameters are fine tuned to realize a yield $>$ 99\% and pull strength $f$ such that the mean $\mu_{f}> 5$g and $\frac{\sigma_{f}}{f}< 20$\%, where $\sigma_f$ is the standard deviation in $f$, as shown in Fig.~\ref{Fig:pull}.

\begin{figure}
\centering
\includegraphics[width=0.8\columnwidth]{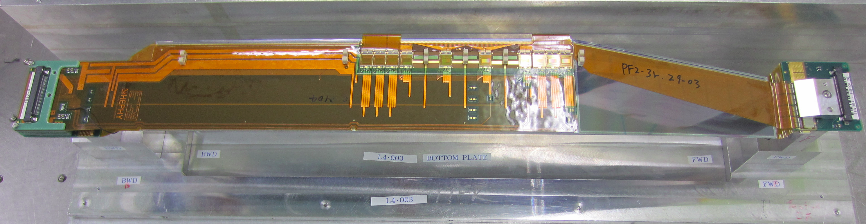}
\caption{A completed layer 4 ladder.} \label{Fig:L4}
\end{figure}

Thermally insulating Airex sheets (light-weight styrofoam) are placed between the DSSD and the readout circuits to minimize the heat transfer between them. This also provides electrical isolation and hence avoids the signal cross-talk. The APV25 chips on the origami flexible circuits are thinned down to 100 $\mu$m to further reduce the material budget. Each ladder is supported by ribs built from carbon-fiber reinforced Airex foam, which is very light but strong and stiff.  A completed L4 ladder is shown in Fig.~\ref{Fig:L4}.

A major challenge during the mass production was that the glue joint between the flexible circuit and the forward DSSD was found to be lifting off on some L4 and L6 ladders. The likely cause was the small overlap between the sensor and the circuit and the stress due to the bending angle. A glue reinforcement strategy is implemented to tackle this issue. 

\section{Quality assurance}

\begin{figure}
\centering
\includegraphics[width=0.4\columnwidth]{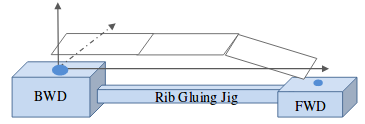}
\caption{Ladder coordinate frame for L4.}\label{Fig:coord}
\end{figure}

The geometrical and electrical quality of the produced ladders are rigorously tested at the assembly sites as well as at KEK, where the ladders are mounted to the support structure. The geometrical precision is measured with an optical Coordinate Measuring Machine (CMM). The position of each sensor is measured and the deviations from the designed values are calculated. The coordinate system used is shown in Fig.~\ref{Fig:coord}. 

Deviation upto 150 $\mu$m in the x-y plane and 200 $\mu$m along the z axis are allowed. The deviations are found to be within the tolerance limits for all the ladders in all the layers. An example of CMM results for an L4 ladder is given in table~\ref{cmm-l4}.

\begin{table} [ht!] 
\centering
\begin{tabular} {c c c c }
\hline 
  & FW & CE & BW   \\[0.5ex]
\hline
\hline
$\Delta$x ($\mu$m) &  $-$44 & ~~6 & $-$47\\[0.5ex]
$\Delta$y ($\mu$m) & ~~25 & $-$21 & $-$9\\[0.5ex]
$\Delta$z ($\mu$m) & ~~174 & $-$139 & $-$81\\[0.5ex]
slant angle ($^{\circ}$) & $-$11.91 & $-$0.07 & $-$0.03\\[0.5ex]
tilt angle ($^{\circ}$) & ~~0.09 & ~~0.09 & ~~0.02 \\[0.5ex]

\hline
\end{tabular}
\caption{Typical CMM results for an L4 ladder. FW, CE and BW stands for forward, central and backward DSSDs.}\label{cmm-l4}
\end{table}

\begin{figure}
\centering
\includegraphics[width=0.6\columnwidth]{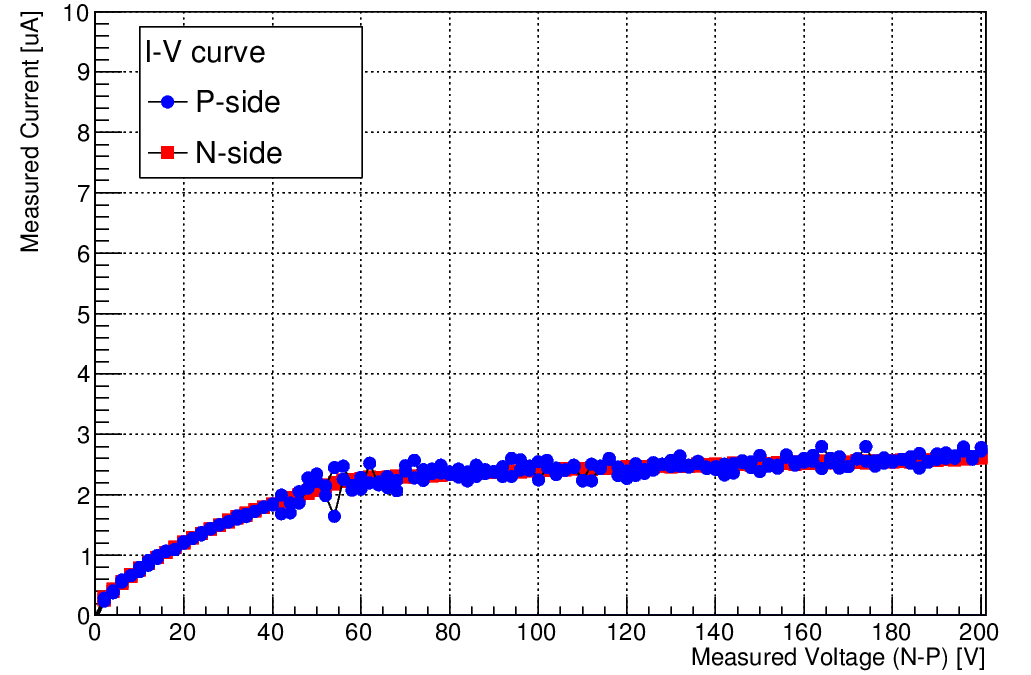}
\caption{I-V characteristics for an L4 DSSD.}\label{Fig:iv}
\end{figure}

Electrical quality of the connections between the sensor and the readout chips is tested. Electrical signals are randomly triggered to evaluate noise, raw noise and the pedestal for each channel. The I-V characteristics of each sensor are analysed and a typical plot is given in Fig.~\ref{Fig:iv}. A source scan is performed using a $\beta$ source ($^{90}$Sr). All ladders have been tested to have good electrical response.

\begin{figure}
\centering
\includegraphics[width=0.5\columnwidth]{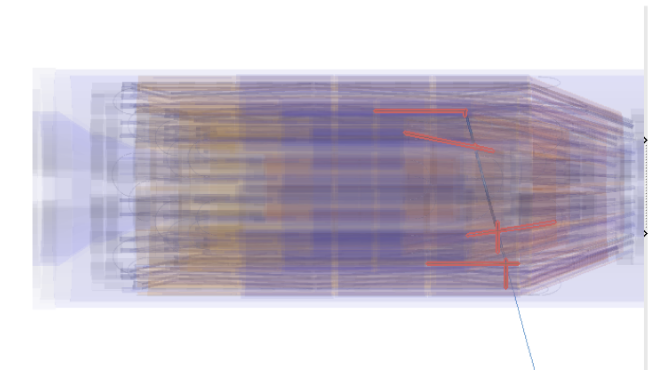}
\caption{First cosmic track seen in the full SVD on July 17, 2018.}\label{Fig:fullsvd}
\end{figure}

A module with one ladder from each layer was tested at DESY beam line in April 2016~\cite{beamtest}. Excellent strip hit efficiency of $>$ 99\% was obtained. This module is also tested during the pilot run of Belle II during April\textendash July 2018. The obtained signal to noise ratio and hit time resolutions are in agreement with the Monte Carlo expectations. All the SVD ladders have been mounted to the final structure successfully and an event display of the first track, induced by a cosmic ray muon, is shown in Fig~\ref{Fig:fullsvd}. 

\section{Conclusions}

The SVD plays a crucial role in Belle~II obtaining its physics goals by providing excellent spatial resolution. The assembly procedure for the ladder modules has been established after rigorous research. All the ladders have been produced with good electrical and mechanical quality. They have been mounted to the final structures and are ready for the first Physics run, which is expected to kick-off in early 2019.

\acknowledgments
I would like to thank the organisers and IIT Madras for successfully being able to attend this conference. This work is supported by MEXT, WPI, and JSPS (Japan); ARC (Australia); BMWFW (Austria); MSMT (Czechia); AIDA-2020 (Germany); DAE and DST (India); INFN (Italy); NRF-2016K1A3A7A09005605 and RSRI (Korea); and MNiSW (Poland).

\end{document}